# AXION SEARCHES AT CERN WITH THE CAST TELESCOPE

C. ELEFTHERIADIS [i*], S. ANDRIAMONJE [b], E. ARIK [c], D. AUTIERO [a],
F. AVIGNONE [d], K. BARTH [a], E. BINGOL [m], H. BRAUNINGER [e],
R. BRODZINSKI [f], J. CARMONA [g], E. CHESI [a,p], S. CEBRIAN [g], S. CETIN [c]
G. CIPOLLA [a], J. COLLAR [h], R. CRESWICK [d], T. DAFNI [m], M. DAVENPORT [a],
R. DE OLIVEIRA [a], S. DEDOUSSIS [i], A. DELBART [b], L. DI LELLA [a],
G. FANOURAKIS [j], H. FARACH [d], H. FISCHER [k], F. FORMENTI [a],
T. GERALIS [j], I. GIOMATARIS [b], S. GNINENKO [l], N. GOLOUBEV [l],
R. HARTMANN [e], M. HASINOFF [a,p], D. HOFFMANN [m], I. G. IRASTORZA [a],
J. JACOBY [m], D. KANG [k], K. KOENIGSMANN [k], R. KOTTHAUS [n], M.KRCMAR [o],
M. KUSTER [e], B. LAKIC [o], A. LIOLIOS [i], A. LJUBICIC [o], G. LUTZ [n],
G. LUZON [g], H. MILEY [f], A. MORALES [g], J. MORALES [g], M. MUTTERER [m],
A. NIKOLAIDIS [i], A. ORTIZ [g], T. PAPAEVANGELOU [i], A. PLACCI [a],
G. RAFFELT [n], H. RIEGE [a], M. SARSA [g], I. SAVVIDIS [i], R. SCHOPPER [m],
I. SEMERTZIDIS [m], C. SPANO [k], V. VASILEIOU [i], J. VILLAR [g], B. VULLIERME [a],
L. WALCKIERS [a], K. ZACHARIADOU [j], K. ZIOUTAS [a,i]

[a] *European Organization for Nuclear Research (CERN), Geneve, Switzerland*
[b] *DAPNIA, Centre d'Etudes de Saclay (CEA-Saclay), Gif-Sur-Yvette, France*
[c] *Department of Physics, Bogazici University, Istambul, Turkey*
[d] *Department of Physics and Astronomy, U. South Carolina, Columbia, Sc, USA*
[e] *Max-Planck-Institut fur Extraterrestrische Physik, MPG, Garching, Germany*
[f] *Pacific Northwest National Laboratory, Richland, Wa, USA*
[g] *Instituto de Fisica Nuclear y Altas Energias, Universidad de Zaragoza, Zaragoza, Spain*
[h] *Enrico Fermi Institute, University of Chicago, Chicago, Il, USA*
[i] *Aristotle University of Thessaloniki, Thessaloniki, Greece*
[j] *National Center for Scientific Research "Demokritos" (NRCPS), Athens, Greece*
[k] *Albert-Ludwigs-Universit. at Freiburg, Freiburg, Germany*
[l] *Institute for Nuclear Research (INR), Russian Academy of Sciences, Moscow, Russia*
[m] *Institut fur Kernphysik, Technische Universitat Darmstadt, Darmstadt, Germany*
[n] *Max-Planck-Institut f.ur Physik, Munich, Germany*
[o] *Ruder Boskovic Institute, Zagreb, Croatia*
[p] *Department of Physics and Astronomy, U. of British Columbia, Vancouver, Canada*

The CERN Axion Solar Telescope (CAST) searches for axions coming from photon to axion conversion in the sun's core, as stated by the Primakoff effect. Axions arise in particle physics as a consequence of the breaking of Peccei – Quinn symmetry which has been introduced as a solution to the strong CP problem. As cosmological axions they are candidates for at least some part of cold Dark Matter.

[*] attending speaker, email: christos.eleftheriadis@cern.ch

They are also expected to be produced copiously in stellar interiors with energies as high as the thermal photons undergoing photon to axion conversion. In our sun the axion energy spectrum peaks at about 4.4 keV, extending up to 10 keV. CAST collected preliminary data in 2002 and data taking with its full capability will start in the beginning of 2003.

## 1. Introduction

Axions constitute a possible candidate for dark matter. Since their appearance in particle physics as a solution to the strong CP problem [1], a number of searches [2, 3, 4, 5] have been reported setting limits to parameters such as the axion-photon coupling. Such limits are also obtained from astrophysical considerations [6, 7]. Axions are expected to be produced in stellar interiors, via the Primakoff effect (see Figure 1) in the scattering of thermal photons off nuclei. For our sun, theoretical expectations are for an axion emission spectrum in the range of a few keV [6, 8]. For supernova explosions this energy range is shifted considerably higher, in the range of 160 MeV. The CAST telescope at CERN makes use of inverse Primakoff effect, which is axion conversion to a photon carrying its energy and momentum, under the influence of a strong magnetic field acting as a catalyst [9]. The potential of the CAST experiment in this search for exotic particles sets much higher standards than any previous axion detector as well as the up to now astrophysical considerations.

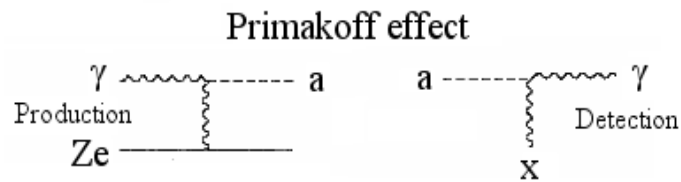

Figure 1. Photon to axion and axion to photon conversion through the Primakoff effect.

## 2. Experimental program

The main component of CAST is a 10 meters long, twin aperture decommissioned LHC prototype magnet, reaching a field of 9 Tesla. Axions produced in the sun's core via the Primakoff effect (5, 6), escape the sun since their coupling to matter is very weak and may enter to the magnet when it is pointing to the sun. Under the influence of the strong transverse magnetic field (B), the axion traversing the 10 meter long (L) magnet has a probability to be transformed to a photon. This is a coherent phenomenon, the probability being proportional to $B^2L^2$, as given by:

$$P_{ag} = 2.1 \times 10^{-17} \left(\frac{B}{9T}\right)^2 \left(\frac{L}{10m}\right)^2 (g_{agg} \times 10^{10} GeV^{-1})^2 |M|^2 \qquad (1)$$

the matrix element $|M|^2=2(1-\cos qL)/(qL)^2$ being equal to unity as far as the coherence holds; $q$ stands for momentum exchange and $g_{a\gamma\gamma}$ for axion – photon coupling. The converted photon carries the energy and momentum of the axion. In terms of $B^2L^2$, the potential of CAST as an axion to photon converter is some 100 times higher compared to the best already existing facility, operating in the Tokyo University [10].

The coherence holds for low axion masses, less than $10^{-2}$ eV. For higher axion masses, the coherence is lost and the axion to photon conversion probability diminishes. CAST plans to explore the range of higher axion masses by filling the magnet tube with helium gas, giving an effective mass to the photon and restoring the coherence. However, the addition of the gas, apart from the restoration of coherence, has other consequences, such as the absorption of photons, which prevent the use of high pressure gas. In fig.2 the exclusion limits are shown for various information sources.

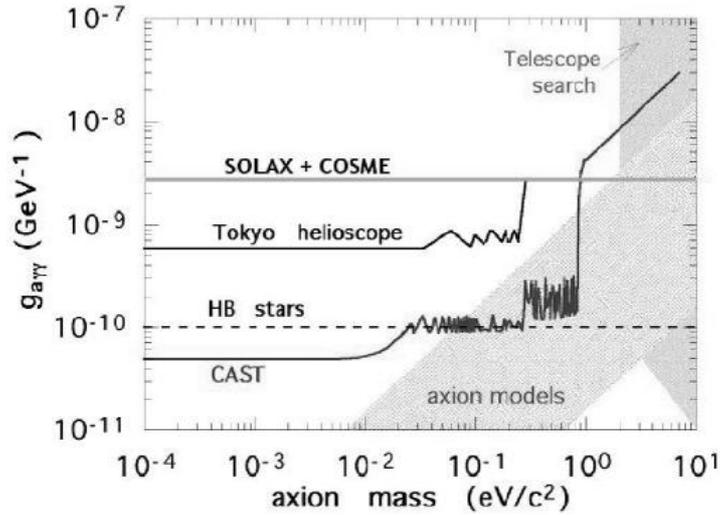

Figure 2. Exclusion plot for CAST, as well as for the SOLAX, COSME and Tokyo experiments. The bound imposed by Horizontal Branch stars and the corridor favored by theoretical axion models are also shown.

It is evident from the exclusion plot in fig. 2 that stringent constraints to axion-photon coupling are provided from astrophysical considerations, based on cooling rates of stars [7,11]. Other limits are provided from experimental searches [3,4,10].

Combining experimental results with astrophysical considerations about cooling rates of stars and cosmological arguments [6,7,9], it seems that there is a window for the axion mass in the range

$10^{-5}$ eV $< m_a < 10^{-3}$ eV.

Considerations for axions with zero coupling to leptons permit higher axion masses, which are otherwise excluded. For instance, the white dwarf cooling time scale excludes lepton coupled axions in the mass range

0.03 eV $< m_a <$ 10 keV.

However, there is still an open possibility that axions do not couple to leptons, according to the so-called hadronic axion models. In that case the above limit does not apply.

### 3. Experimental set up

The magnet is placed on a moving platform and can be moved up to $\pm 8^0$ vertically and $\pm 40^0$ horizontally. The magnet can track the sun 1.5 hours during sunrise and 1.5 hours during sunset. Data taking during the remaining hours of the day will be reference background measurements.

Three detectors have been developed for the CAST experiment, namely a CCD, a TPC and a micromegas [7]. The TPC looks into the two openings of one side of the magnet. Special care was taken to reduce as much as possible the noise in all detectors. In order to improve further the signal to noise ratio, an X ray focusing device will be used, which has been constructed for the ABRIXAS X ray space mission. With this configuration, X rays from axion to photon conversion within the magnet are focused on a millimeter area of the detector improving significantly the signal to noise ratio. The X ray focusing system has been simulated [12] and the results were in excellent agreement with test data taken in Max Planck Institute in Munich.

## 4. Conclusions

During the first year of full data taking, CAST will cover an axion rest mass below ~0.01 eV, whereas later data will be taken with He gas in order to search for axions with rest mass up to ~1 eV. The magnet will track the sun for some 200 minutes per day, whereas the remaining time will be devoted to background measurements. The large BL value of the CAST magnet along with the use of the focusing X-ray telescope, distinguish CAST's performance from the one of previous experiments. Data taken with the X-ray focusing device provide a simultaneous background measurement from the active area of the detector around the ~1 mm focal spot.

There are still unexplored possibilities looking at the sun itself [13]. In addition, it is possible, during normal operation of CAST, to search for a signature from other astrophysical sources, such as the galactic center, supernovae or GRBs, which will be within the field-of-view of CAST. Axions or axion – like particles with possibly much higher energy will need appropriate detectors. Data expected from CAST during the next three years, as well as parallel theoretical work, might provide a new input for this field of astroparticle physics.